\RequirePackage{fix-cm}
\documentclass[smallextended]{svjour3}       % onecolumn (second format)
\smartqed  % flush right qed marks, e.g. at end of proof
\usepackage{graphicx}
\usepackage{url}
\usepackage{cite}
\usepackage{amsmath,amssymb,amsfonts}
\usepackage{algorithmic}
\usepackage{graphicx}
\usepackage{textcomp}
\usepackage{xcolor}
\usepackage{booktabs}
\usepackage{multirow}
\usepackage{hyperref}

\usepackage{makecell}
\usepackage{subfig}
\usepackage{natbib}
\usepackage{tcolorbox}
\usepackage{subcaption}

\begin{document}

\title{COBOLAssist: Analyzing and Fixing Compilation Errors for LLM-Powered COBOL Code Generation}
% \subtitle{Do you have a subtitle?\\ If so, write it here}


\author{Anh T. V. Dau         \and
        Shin Hwei Tan  \and
        Jinqiu Yang \and
        Nghi D. Q. Bui \and
        Anh Tuan Nguyen
        %etc.
}


\institute{Anh T. V. Dau \at
              Concordia University, Montreal, Canada \\
              \email{thivananh.dau@mail.concordia.ca}           %  \\
           \and
           Shin Hwei Tan \at
              Concordia University, Montreal, Canada \\
              \email{shinhwei.tan@concordia.ca} 
            \and
            Jinqiu Yang \at
              Concordia University, Montreal, Canada \\
              \email{jinqiu.yang@concordia.ca} 
            \and
            Nghi D. Q. Bui \at
              FPT Software AI Center, Vietnam \\
              \email{nghibdq@fpt.com} 
            \and
            Anh Tuan Nguyen \at
              FPT Software AI Center, Vietnam \\
              \email{anhnt446@fpt.com} 
}

\date{Received: date / Accepted: date}

\maketitle

\begin{abstract}
Legacy programming languages such as COBOL (Common Business-Oriented Language) remain critical in business computing. However, maintaining legacy COBOL systems is increasingly challenging due to a declining pool of skilled developers and the persistence of COBOL errors that require deep domain expertise to resolve. This paper investigates the challenges of COBOL compilation errors and introduces a framework leveraging large language models (LLMs) to address these issues. 
We first categorize the common compilation errors in LLM-generated COBOL code into three groups: incomplete code errors, syntax errors, and type-related errors. We further propose COBOLAssist, a technique to enhance code correctness through iterative repairs guided by compilation feedback. Our evaluation using five LLMs including GPT variants and mAInframer, shows a high prevalence of incorrect program structures and function usage in COBOL programs and demonstrates the effectiveness of COBOLAssist, with the compilation success rates increasing from 29.5\% to 64.38\% for GPT-4o-mini and from 41.8\% to 95.89\% for GPT-4o. It also improves pass@1 significantly, for example from 9.1 to 22.6 for GPT-4. Notably, while mAInframer-34B achieves the highest compilation success rate, its functional correctness remains limited. This research not only highlights the limitations in current LLMs for COBOL but also demonstrates a practical path forward for automated debugging in legacy systems.

% By categorizing the common types of compilation errors that arise in LLM-generated COBOL code and evaluating the effectiveness of our proposed method through experiments with state-of-the-art models such as GPT-4o-mini, GPT-4o, and mAInframer, we aim to empower developers with advanced tools for error resolution. Our findings reveal significant improvements in compile success rates and correctness after the application of the self-debugging technique, thereby demonstrating its potential to enhance the quality and reliability of automatically generated COBOL code. This research not only provides insights into the challenges faced in COBOL compilation but also offers automated solutions that can bridge the skill gap in maintaining legacy systems.
\end{abstract}

\section{Introduction}
COBOL systems have played a fundamental role in business operations since their inception in the late 1950s. Designed primarily to handle large-scale data processing, COBOL became the backbone of government and corporate computing systems worldwide. Despite the rapid emergence of modern programming languages, COBOL remains an integral part of global infrastructure, supporting mission-critical applications such as banking transactions, insurance processing, and government record keeping \citep{ciborowska2021contemporary,upadhaya2023understanding}. It is estimated that there are currently approximately 220 billion lines of COBOL code in use, with 1.5 billion lines being written each year \citep{taulli_cobol_nodate}, indicating the importance of COBOL code maintenance. %Due to its dependability and effectiveness, COBOL is ensuring that vital services will continue to run efficiently even as the underlying technology ages.

Maintaining COBOL applications presents several challenges, particularly due to the diminishing number of skilled COBOL developers. As experienced programmers retire, the knowledge required to debug, optimize, and enhance legacy COBOL code becomes harder to acquire. According to a report by American Banker, banks struggle to recruit young engineers with COBOL skills \citep{noauthor_wanted_2014}. Furthermore, many COBOL systems were written decades ago with minimal documentation, making debugging a time-consuming and complex process \citep{gaie2021cost}. 
%\jinqiu{Revise the logic below, no need to go to the challenges of COBOL. Just write about "with this rise of LLM-powerd code generation, it is great potential to use LLM to ease the faced challenges of COBOL development. However, COBOL is a low-resource language based on open-source repos etc. so LLM may not have sufficient trained on COBOL...''}
With the rise of LLM-powered code generation, there is growing potential to use these models to ease the challenges faced in COBOL development and maintenance~\citep{dau2024xmainframe,gandhi2024translation,noauthor_bloop_nodate}.

However, COBOL is a low-resource language with only a limited number of open-source repositories and public datasets available, which means that there exists only a limited corpus to train LLMs specifically for COBOL. This causes LLMs' failure to handle COBOL-specific syntax and semantics effectively, resulting in low compilation success rates and frequent generation errors \citep{noauthor_bloop_nodate,dau2024xmainframe,gandhi2024translation}. 
Although LLMs have exhibited great potential to repair compilation errors and software defects in many programming languages such as Python, Rust, C, etc. \citep{zhang2024pydex,deligiannis2024rustassistant,sevenhuijsen2024vecogen}, similar efforts are limited in adopting LLMs for COBOL, especially in fixing compilation errors and defects. A prior study from BloopAI \citep{noauthor_bloop_nodate} benchmarks LLM performance on COBOL generation but does not evaluate automated error correction tasks (e.g., fixing compilation errors and defects in COBOL code). Although trained on vast amounts of natural language and code data, LLMs, such as GPT-4 \citep{gpt4} and CodeLlama, \citep{roziere2023code} do not perform well in COBOL code generation\citep{noauthor_bloop_nodate,dau2024xmainframe}.
This gap motivates our work in enhancing LLM-based techniques to improve the reliability of generated COBOL code through automated compilation error repair.
To improve the reliability of COBOL code generation, we conduct the first study of the types of COBOL compilation errors in LLM-generated code,
which exhibit different patterns compared to the compilation errors induced by human developers.
% Compilation errors present a significant challenge in maintaining legacy systems, as they often require manual intervention and deep domain expertise. While human developers commonly encounter these errors, LLMs generate additional error patterns that may not typically arise in manually written code. 
%\jinqiu{Simply the text above. Just say something ``a lack of understanding on the compilation errors by LLM for cobol gen, which may exhibit different patterns compared to the compilation errors induced by human developers.}
%To compare the characteristics of LLM-generated COBOL code, 
Specifically, we designed a compilation error categorization tailored for COBOL code, consolidating and refining existing error categorization from a prior study of COBOL compilation errors in human-written code \citep{litecky1974study}. %and real-world datasets.
Our study identifies three categories of errors in LLM-generated COBOL code: (1) incomplete code errors, (2) syntax errors, and (3) type-related errors. 
% \jinqiu{Coming back on this.}

Motivated by our study that highlights the prevalence of compilation errors in LLM-generated COBOL code, we introduce COBOLAssist, a self-debugging approach that enables LLMs to iteratively correct compilation errors in COBOL code without external intervention. 
% Specifically, we propose a structured prompting technique tailored for fixing COBOL compilation errors and\jinqiu{below is part of evaluation no? like ablation analysis. need to re-structure the writing here.} systematically assess the impact of each prompt component on debugging performance.
To evaluate our approach, we conduct an empirical study on the COBOLEval benchmark \citep{noauthor_bloop_nodate} with five models, including GPT variants (including GPT-3.5-Turbo (GPT-3.5 in short), GPT-4, GPT-4o-mini, and GPT-4o \citep{gpt3.5,achiam2023gpt,gpt4o}), and mAInframer \citep{noauthor_bloop_nodate}, the only open-source fine-tuned LLM for COBOL generation. Our results demonstrate that COBOLAssist consistently improves the syntactic correctness of generated programs across all models, substantially reducing compilation failures.
%\jinqiu{We applied [tool name] to repair ...}
%\jinqiu{need to put concrete results in. how the compilation errors are different from human written ones? what is the result of the tool how effective it is.}
% Then, we applied COBOLAssist to fix the compilation errors generated by these models. Starting from an LLM-generated program, COBOLAssist leverages compiler messages to guide the code modifications until the generated program compiles successfully or a stopping condition is reached. 

Our study answers three research questions:
% \begin{itemize}

\noindent \textbf{RQ1: What are the major categories of compilation errors in LLM-generated COBOL code?} We systematically categorize the compilation errors produced by LLMs and compare them with those found in human-written COBOL programs \citep{litecky1974study}. Our analysis reveals that 35.1\% of errors are due to incorrect use of  program structures—almost double the rate in human code (19.8\%). Furthermore, two error types, \textit{Incorrect Use of Function} (17.2\%) and \textit{Incomplete Block Termination} (5.6\%), are unique to LLM-generated code, highlighting gaps in the model's understanding of COBOL’s semantics.  

\noindent \textbf{RQ2: How effective is COBOLAssist in resolving COBOL compilation errors?} COBOLAssist significantly boosts compilation success across all models. For instance, GPT-4o’s compilation success rate increases from 41.8\% to 95.89\%, and mAInframer-34B reaches 97.94\%. COBOLAssist also improves semantic correctness: GPT-4's pass@1 rises from 9.1 to 22.6, and GPT-4o from 16.4 to 29.45.

\noindent \textbf{RQ3: What types of compilation errors remain unresolved after using COBOLAssist?}
Despite the improvements, some error types remain difficult to fix automatically. Our analysis shows that unresolved cases often involve semantic misunderstandings or complex structural inconsistencies that exceed the model's reasoning capabilities, pointing to the need for deeper program understanding or hybrid AI-human debugging.

In summary, our work makes the following contributions:
% \href{https://zenodo.org/records/15386580?token=eyJhbGciOiJIUzUxMiJ9.eyJpZCI6ImQ1OTQyYWJlLTcxYmQtNDMzMS04ZWE1LWEzMzc2OWE4NGYxYyIsImRhdGEiOnt9LCJyYW5kb20iOiI3MmMzNTc5ZDUyMGFmMTM1Mjc3Yjc2ODcyNDFiMzBlMiJ9.Cvd0a1_p-JxfECciqr6zsq10hKTbT4oIue2fib7iKnbf0vkTw6oyU_Any7XkIdjRX23ThK8PkFgc2BmM3lEs8Q}{link}}
\begin{itemize}
\item We present the first  empirical study of compilation errors in LLM-generated COBOL programs and contrast them with errors observed in human-written code to highlight fundamental differences in error patterns.

\item We introduce COBOLAssist, a compiler-guided self-debugging framework that iteratively leverages compiler feedback to improve the syntactic correctness of LLM-generated COBOL code.

\item Through an evaluation on 146 COBOL programming tasks across multiple LLMs, we show that our technique substantially improves the compilation success rate and functional correctness for LLM-generated code.

% \item We analyze the types of errors that remain unresolved after using COBOLAssist to provide guidance for future repair strategies.

\item We provide an analysis of unresolved and incorrectly repaired cases to identify common failures and shed light to guide future research on robust program repair for legacy languages.

% \item We provide insights into common failure modes, guiding future research in enhancing LLM-driven generating and automated debugging for COBOL.
\end{itemize}

\section{Background}

\subsection{COBOL (Common Business Oriented Language)}

COBOL is one of the earliest high-level programming languages and designed with the goal of making programming as close to English prose as possible. As a result, COBOL programs are highly verbose and rely on structural concepts drawn from natural language to organize code. Its list of reserved words is extensive, containing hundreds of keywords that enhance readability but also increase syntactic complexity.

\begin{figure}[h]
\centering
\includegraphics[width=0.45\textwidth]{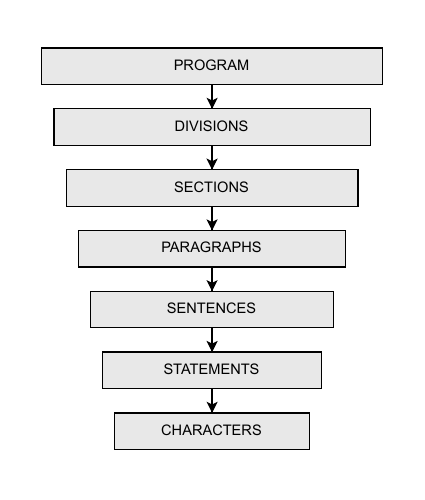}
% \caption{Structure of COBOL program.}\footnote{This figure is from \url{https://www.tutorialspoint.com/cobol/cobol_program_structure.htm}}
   \caption[Caption for LOF]{Structure of a COBOL program.\footnotemark}
  % \small\textsuperscript{} This figure is from \url{https://www.tutorialspoint.com/cobol/cobol_program_structure.htm}
\label{fig:structure_cobol}
\end{figure}
\footnotetext{This figure is adapted from \url{https://www.tutorialspoint.com/cobol/cobol_program_structure.htm}}

\begin{figure}
\centering
\includegraphics[width=0.65\textwidth]{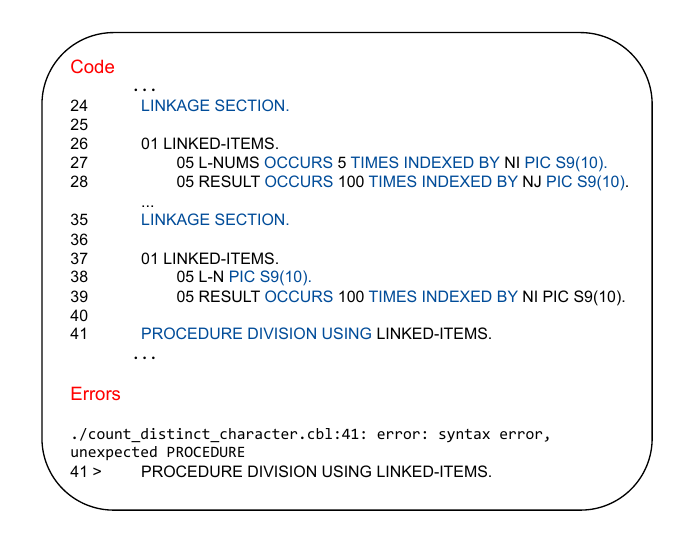}
\caption{A COBOL program with two \texttt{LINKAGE SECTION} declarations, which violates COBOL's structural rules, leading to a syntax error at line 41.}
\label{fig:incorrect_structure}
\end{figure}

 \noindent\textbf{COBOL syntax.} COBOL uses a structured programming approach \citep{tompkins1983defense} where programs are organized by sections and divisions, as shown in Figure 
 \ref{fig:structure_cobol}. 
 A COBOL program is organized into four main divisions, appearing in a fixed order:
(1) Identification Division, (2) Environment Division, (3) Data Division, and (4) Procedure Division. The \texttt{IDENTIFICATION DIVISION} provides metadata about the program for both the programmer and the compiler. The \texttt{ENVIRONMENT DIVISION} specifies the system context in which the program is expected to run. The \texttt{DATA DIVISION} defines all the variables and data structures used throughout the program. Finally, the \texttt{PROCEDURE DIVISION} contains the executable logic, including the algorithm that manipulates the defined data using control flow constructs (e.g., sequence, selection, and iteration). 
The two most important divisions are \texttt{DATA DIVISION} and \texttt{PROCEDURE DIVISION}, where data definitions and executable logic reside, respectively.
% The data division of COBOL has three sections: (1) \texttt{FILE SECTION} (i.e., describe the data that is sent to or coming from the computer’s peripherals), (2) \texttt{WORKING-STORAGE SECTION} (i.e., describes the program variables), and (3) \texttt{LINKAGE SECTION} (i.e., establishes a link between programs). 
\texttt{DATA DIVISION} is typically divided into two primary sections: \texttt{FILE SECTION}, which defines input and output data associated with external devices such as files or terminals, and \texttt{WORKING-STORAGE SECTION}, which declares general-purpose variables used throughout the program. In some programs, a third section, \texttt{LINKAGE SECTION}, is used to pass data between different programs or subprograms.  Figure \ref{fig:incorrect_structure} shows an incorrect LLM-generated program with duplicated linkage sections. 

\texttt{PROCEDURE DIVISION} operates on the data defined in the \texttt{DATA DIVISION} and is organized hierarchically into Sections, Paragraphs, Sentences, and Statements. While sections are optional, every \texttt{PROCEDURE DIVISION} must contain at least one paragraph, sentence, and statement. Paragraph and section names are user-defined and should reflect the specific operations being performed. This structure supports a clear, modular algorithm design but also increases the syntactic complexity, which can be challenging for automatic code generation systems to handle correctly.

\subsection{Code Generation using Large Language Models}
% \jinqiu{This is too general, revise it to show the common workflow of LLM for code generation, there is a prompt, generated code, then either tools or human trying to fix the errors.}
% \fixed{
Large Language Models (LLMs) have shown remarkable capabilities in code generation across a variety of programming languages \citep{hui2024qwen2,roziere2023code,achiam2023gpt}. However, the generated code is often imperfect. It may suffer from syntax errors, semantic bugs, or compilation failures \citep{pan2024lost,sevenhuijsen2024vecogen}. Despite growing interest in LLM-based repair and refinement, most prior research has focused on popular languages like Python or Java, with little attention paid to COBOL—a legacy but still critical language. This raises the question: \textit{how well do current LLMs perform in generating COBOL code, especially given their limited training data?} To address this, COBOL-specific models such as mAInframer \citep{noauthor_bloop_nodate} have been developed, aiming to improve code generation for legacy mainframe systems.
While models can be taught to fix general syntax or logic bugs, compilation errors in COBOL pose challenges due to its rigid structure and constraints. 
In this study, we focus on understanding and reducing COBOL compilation errors by integrating LLMs into a feedback-driven workflow, where compiler error messages guide automatic self-debugging.

\section{Methodology}

% \subsection{The Manual Compilation Error Labeling Process}
% \jinqiu{this should be before the categorization results.}
% \jinqiu{OK, so move this to the methodology.}
\subsection{Construction of COBOL Compilation Error Categorization}

While previous work has studied COBOL compiler errors in traditional programming scenarios \citep{litecky1974study} or analyzed general code quality issues in LLM-generated code \citep{tambon2024bugs}, these efforts have not focused on the unique challenges posed by COBOL code generation. To bridge this gap, we develop a tailored categorization scheme for LLM-induced COBOL compilation errors following the three steps below.

\noindent \textbf{Step 1: LLM Code Generation.}
In this step, we first select three recent leading LLMs: GPT-4o, GPT-4o-mini, and mAInframer-34B, to generate COBOL code based on the COBOLEval benchmark \citep{noauthor_bloop_nodate}, which is the only publicly available dataset for this domain and includes 146 COBOL programming tasks. From these models, we collect 876 generated COBOL programs. They are then compiled using the GnuCOBOL compiler \citep{noauthor_gnucobol_nodate}, yielding a total of 980 compilation errors, detailed in Table \ref{tab:avg_round} (due to the significant manual efforts in labeling errors, we do not categorize errors from code generated by GPT-3.5 and GPT-4 shown in this table). 

\noindent \textbf{Step 2: Initial Category Distillation.} 
Two human annotators, who are a researcher and COBOL developer with 5 years of experience, review the generated code and associated compilation errors. Then, they start clustering similar errors and assigning descriptive tags to identify underlying causes. Through discussion and iterative refinement, they create an initial hierarchical taxonomy that captures the key categories of COBOL compilation bugs, including both traditional syntax issues and error types specific to LLM-generated code. 

\noindent \textbf{Step 3: Independent Labeling and Refinement.}
Using the initial taxonomy, each labeler independently annotates the full set of 980 compilation errors. After the independent phase, they compare their annotations to identify mismatches. The Cohen's Kappa co-efficient is 0.9, indicating high agreement between annotators. Disagreements are reviewed and discussed collaboratively to ensure consistent and accurate classification. In cases where existing categories are insufficient to describe the observed error, new categories were introduced through consensus. This iterative process yields the final error categorization used throughout our analysis.

% \todo{Need to describe how the categorization/taxonomy is concluded, normally done by open coding process. You can check my paper on TVM bugs (EMSE, 2023) about how to describe the process.}
% \jinqiu{also, does the prior textbook categorization inspire you in any way? If does, then need to describe it.}
% Our categorization bridges this gap by combining existing knowledge and empirical studies\todo{which existing knowledge which empirical studies? needs to have citation}.

% This process involves two human annotators, who are researchers and COBOL developers with 5 years of experience. It focuses on categorizing errors produced by LLM-generated code, using a pre-defined taxonomy of compilation error categories. Each annotator first independently reviews and assigns category labels to all compilation errors based on the taxonomy. Then, they compare their individual annotations and resolve any discrepancies. Any cases with inconsistent or unclear classifications are discussed collaboratively to determine the most appropriate category. \todo{In total, the annotators categorized a total of .. errors. These errors are generated by ... Due to the significant manual efforts in labeling errors, we only categorizes errors in code generated by .... } \fixed{We describe the details of the dataset containing the compilation errors in Section \ref{rq1_results}.}

\subsection{COBOL Compilation Error Categorization}

Given the complexity of COBOL syntax and structure, we categorize errors into three main types: (1) incomplete code errors, (2) syntax errors, and (3) type-related errors.%, as shown in Table~\ref{tab:compilation}.

\noindent\textit{\textbf{1) Incomplete Code Errors:}} refer to issues arising from the incorrect formation of program blocks or termination symbols.

\noindent\textbf{Incomplete Block Termination}: COBOL requires explicit termination of control structures and programs using specific keywords such as \texttt{END-IF}, \texttt{END-PERFORM}, \texttt{END-EVALUATE}, and \texttt{END PROGRAM}. If these terminators are omitted, the compiler cannot determine where a logical block ends, leading to syntax errors or unexpected program behavior. While traditional studies of COBOL compilation errors rarely report this issue—since human developers are typically aware of COBOL's strict block termination rules—this error frequently arises in LLM-generated code.
% \jinqiu{so references like this indicate that you are actively comparing LLM compilation errors with human-written, I assume from the textbook? if so this process needs to be reflected in the methodology before introducing the categorization/taxonomy} 
We observe that LLMs often generate incomplete control structures, especially when dealing with long or nested code blocks. 
% As shown in Figure \ref{fig:incomplete_block}, the absence of the required terminator, END-IF, results in incomplete code structures, causing parsing issues during compilation. This error type highlights a new failure pattern unique to LLM-generated COBOL programs, reflecting the challenge of maintaining structural consistency in auto-generated code.

% \begin{figure}[h]
% \centering
% \includegraphics[width=0.45\textwidth]{image/missing_terminator.drawio.pdf}
% \caption{An example of Incomplete Block Termination.}
% \label{fig:incomplete_block}
% \end{figure}

\noindent\textbf{Unterminated Statements}: COBOL statements require explicit termination, usually by a period (.) or other delimiters. If a statement lacks proper termination, the compiler may misinterpret subsequent lines, resulting in cascading errors. This issue is especially common in multiline statements where a period is either missing or incorrectly placed, causing compilation failures or unintended execution flow. 
% Figure \ref{fig:unterminated_statement} illustrates this type of error. Although the compiler reports an error at line 36, the actual cause lies in the misplaced period on line 35.

% \begin{figure}[h]
% \centering
% \includegraphics[width=0.45\textwidth]{image/period.drawio.pdf}
% \caption{An example of Unterminated Statements.}
% \label{fig:unterminated_statement}
% \end{figure}

\noindent\textit{\textbf{2) Syntax Errors}} encompass violations of COBOL's rules regarding the arrangement and formatting of code elements.

\noindent\textbf{Incorrect Use of Program Structures}:
COBOL provides structured programming constructs such as \texttt{IF}, \texttt{PERFORM}, and \texttt{EVALUATE} for decision-making and loops. Misplacing these structures, such as nesting loops incorrectly, leads to syntax errors. Figure \ref{fig:incorrect_structure} demonstrates a common structural mistake made by LLMs: duplicating the \texttt{LINKAGE SECTION}. In COBOL, \texttt{LINKAGE SECTION} allows the called program to access variables sent by the caller, and only one \texttt{LINKAGE SECTION} is allowed per program. In this example, the program declares a \texttt{LINKAGE SECTION} at line 24 and again at line 35, each with different data structures. This leads to a syntax error at line 41, where the compiler encounters the \texttt{PROCEDURE DIVISION}.
% Proper indentation and consistent structure are crucial for preventing these issues, as COBOL is highly dependent on precise formatting.

\noindent\textbf{Incorrect Use of Reserved Word}:
% \sh{need to rename this type to something like Incorrect Use of Reserved Word so that it is consistent with other error type. Currently, reserved word is not an error
COBOL has a set of reserved keywords, such as \texttt{MOVE}, \texttt{COMPUTE}, \texttt{DISPLAY}, etc., that cannot be used as variable or procedure names. Using a reserved word incorrectly can cause ambiguous references, leading to compilation errors. 
% The example in Figure \ref{fig:reserved_word} highlights an incorrect use of a COBOL reserved keyword as a variable name, which conflicts with the language’s predefined syntax rules.

% \begin{figure}[h]
% \centering
% \includegraphics[width=0.45\textwidth]{image/Reserved_word.drawio.pdf}
% \caption{An example of Incorrect Use of Reserved Word (i.e, \texttt{SUM}).}
% \label{fig:reserved_word}
% \end{figure}

\begin{figure}
\centering
\includegraphics[width=0.65\textwidth]{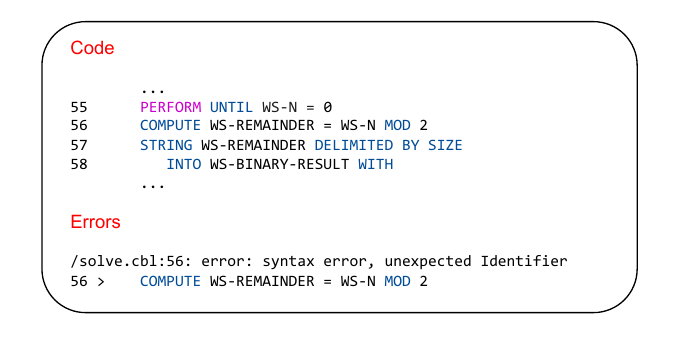}
\caption{An example of Incorrect Use of Built-in Functions. The figure presents a COBOL snippet that incorrectly applies the \texttt{MOD} function as an inline operator rather than using the correct syntax \texttt{FUNCTION MOD(arg1 arg2)}.}
% , \texttt{argument_{2})}}
\label{fig:incorrect_function}
\end{figure}

\noindent\textbf{Incorrect Use of Built-in Functions}:
COBOL includes a set of standardized built-in functions designed to perform common operations, such as mathematical calculations, string manipulation, date handling, etc. These functions must be used with the correct syntax: they are invoked using a specific format—typically \texttt{FUNCTION <name>(arg1 arg2 ...)}. For example, the modulus operation in COBOL should be written as \texttt{FUNCTION MOD(a b)}, where \textit{a, b} are two arguments. Figure \ref{fig:incorrect_function} illustrates an example where LLM used this function incorrectly. The compiler threw a syntax error at line 56, interpreting MOD as an unexpected identifier.
This type of error reflects a semantic misunderstanding by the model, which either hallucinates or misapplies language constructs based on knowledge from other programming languages. These mistakes are more prevalent in LLM-generated COBOL than in human-written code, as prior COBOL usage studies show developers rarely misuse built-in functions \citep{noauthor_bloop_nodate}.

% While prior studies on COBOL compilation errors mainly focused on syntax violations or structural mistakes made by human developers \citep{litecky1974study}, this error type emerges more prominently in LLM-generated code. This is because LLMs often hallucinate or misuse built-in functions, especially when lacking sufficient COBOL-specific knowledge.

\noindent\textbf{Undefined Object}:
% \sh{this error type may need to be changed}}:
This error occurs when a variable, function, section, or paragraph is referenced before being declared. COBOL requires explicit declaration of all objects in the \texttt{DATA DIVISION} or \texttt{PROCEDURE DIVISION}. If a program attempts to use an undefined variable, the compiler generates an error. This is often due to typos, missing declarations, or incorrect scope definitions. 
% Figure \ref{fig:undefined_object} illustrates a scenario where an undefined function leads to reference errors during compilation
% \sh{i think this error type is problematic as you define an object to be variable, section or paragraph in the table but you clearly explained that this is an undefined function?}.

% \begin{figure}[h]
% \centering
% \includegraphics[width=0.45\textwidth]{image/undefined_object.drawio.pdf}
% \caption{An example of Undefined Object}
% % \sh{need to explain what is the undefined object in this example. maybe this example is not correct as the ISPRIME seems to be a function? Should this example fits into incorrect use of function? }.}
% \label{fig:undefined_object}
% \end{figure}

\noindent\textbf{Incorrect Use of Variable}:
COBOL enforces strict typing and scoping rules for variables. Violations such as using variables outside their declared scope, misspelling variable names, or passing incompatible variable types to procedures can lead to compilation errors. While such mistakes are uncommon in human-written COBOL code due to the language's explicit declaration style, they frequently emerge in LLM-generated code. This is largely because LLMs struggle to maintain consistent variable tracking and proper scoping across long and complex code generation. 
% As demonstrated in Figure \ref{fig:incorrect_variable}, an incorrect reference to a variable causes errors during compilation.

% \begin{figure}[h]
% \centering
% \includegraphics[width=0.45\textwidth]{image/incorrect_variable.drawio.pdf}
% \caption{An example of Incorrect Use of Variable\todo{it is unclear which is the incorrect variable here? explain in the text or in this caption}.}
% \label{fig:incorrect_variable}
% \end{figure}

\noindent\textit{\textbf{3) Type-related Errors}} involve discrepancies related to data declarations and type usage within the program.

\noindent\textbf{Incorrect Data Type}: COBOL has strict rules for handling data types, including PIC clauses that define variable formats. Errors occur when there is a type mismatch, such as assigning an alphanumeric value to a numeric field or using an incorrectly formatted COMP (computational) data type. These issues lead to compilation failures and often require explicit type conversions or adjustments in variable definitions. 
% Figure \ref{fig:incorrect_data} is an example showing a data type mismatch, where an operation attempts to assign incompatible data types.

% \begin{figure}[h]
% \centering
% \includegraphics[width=0.45\textwidth]{image/data_type.drawio.pdf}
% \caption{An example of Incorrect Data Type.}
% \label{fig:incorrect_data}
% \end{figure}

\subsection{COBOLAssist}
\begin{figure*}[]
\centering
\includegraphics[width=\textwidth]{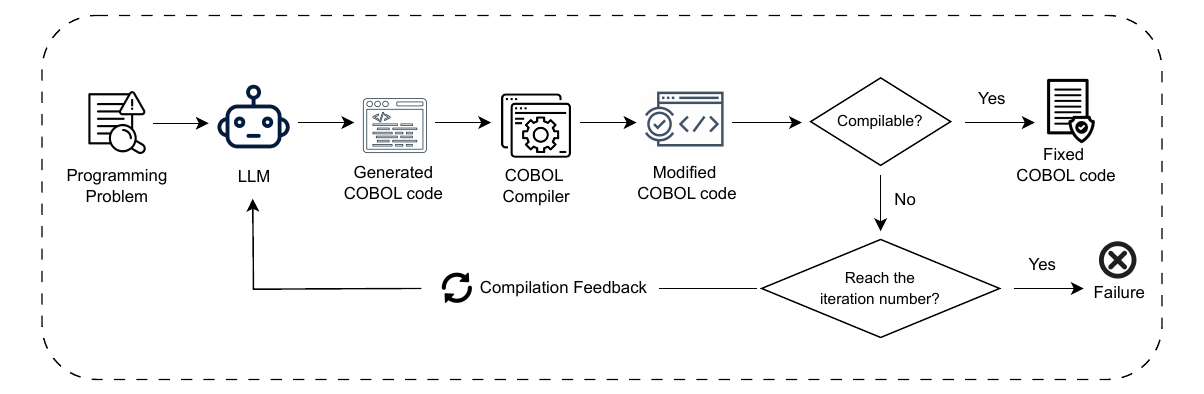}
\caption{Overview of COBOLAssist—a self-debugging technique with compilation feedback.}
\label{fig:overview}
\end{figure*}

% \jinqiu{This methodology is way too short. You need to include a prompt design at least.}
COBOLAssist is designed to improve the correctness of LLM-generated COBOL programs by incorporating compiler feedback into an iterative self-debugging loop. Figure \ref{fig:overview} shows the overall workflow of COBOLAssist. Given a programming problem written in natural language that includes the task description and a template structure, COBOLAssist uses an LLM to generate an initial COBOL code.
% LLMs are used to generate an initial COBOL program corresponding to the given problem. 
Once the initial code is generated, it is compiled using the GnuCOBOL compiler.
If compilation fails, the compiler emits messages, which serve as feedback for triggering the self-repair process. These error messages are then embedded into a follow-up prompt that asks the LLM to revise its previously generated code. 
For optimal results, we design the prompt based on OpenAI's six strategies for achieving better results~\citep{rules}. Specifically, our prompt instructs the model to adapt a persona of an experienced COBOL software engineer and uses section titles (e.g., "COBOL Code" and "Compiler Error Log") to clearly state the distinct parts of the input. 
The prompt used in COBOLAssist is shown below: 

\begin{tcolorbox}
You are an experienced COBOL software engineer with deep knowledge of COBOL syntax, structure, and best practices. Your task is to perform debugging on a given COBOL program with the compilation errors.

Below is the original COBOL code followed by the compiler’s error log. Your job is to revise the code to resolve all compilation errors, ensuring that the corrected program is not only syntactically valid but also logically sound.

Please carefully analyze the error messages and update the code accordingly. Prioritize clarity, maintainability, and adherence to COBOL’s structural rules.

Input:

COBOL Code:
\textit{[code]}

Compiler Error Log:
\textit{[error\_log]}

\end{tcolorbox}
% Instead of relying solely on the model’s internal reasoning, we utilize this compilation feedback to guide the debugging process
This feedback-driven loop continues iteratively. In each iteration, LLMs receive the updated error diagnostics and generate a new code revision. 
The newly modified code is recompiled, and any remaining errors are again passed back to the model. 
% This iterative correction process continues until the program compiles successfully or a predefined stopping criterion is met.
This self-debugging cycle terminates when either (1) the program compiles successfully or (2) a predefined maximum number of iterations is reached.

By integrating compiler feedback into the debugging workflow, COBOLAssist helps LLMs resolve errors they might have on the first attempt. It is particularly useful in addressing structural errors, type mismatches, and incomplete statements, which are otherwise challenging for one-shot code generation.
% this technique enhances LLMs' ability to identify and rectify structural, syntactic, and type-related errors in COBOL programs, ultimately improving the quality of automatically generated code.
% \jinqiu{When will the iterative process terminate? need to describe.}

\section{Evaluation}
\subsection{Experimental Setup}
\begin{figure}[]
\centering
\includegraphics[width=0.45\textwidth]{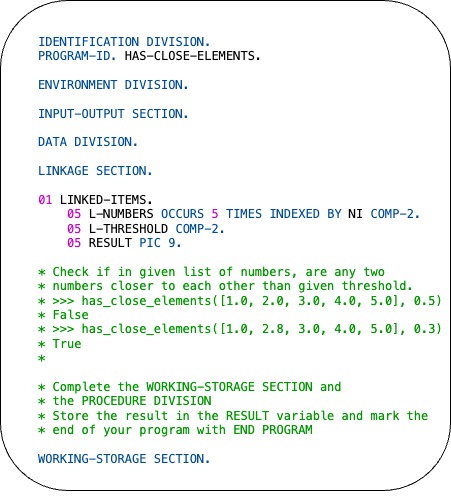}
\caption{An input instance from the COBOLEval benchmark.}
\label{fig:coboleval}
\end{figure}

\noindent\textbf{COBOL code generation benchmark:}
We evaluate the effectiveness of code generation using the only existing COBOL code generation dataset-COBOLEval \citep{noauthor_bloop_nodate}. COBOLEval is the first benchmark designed to evaluate LLM performance in COBOL code generation. It consists of 146 programming problems translated manually from the widely used HumanEval-Python benchmark \citep{chen2021codex}. Each problem includes an average of six test cases, which requires models to generate functionally correct COBOL programs that pass all tests. Figure 
 \ref{fig:coboleval} shows a problem example of COBOLEval. 
 
\noindent\textbf{Evaluated LLMs:}
% \jinqiu{This section can be much shortened.}}
To evaluate the effectiveness of COBOLAssist in fixing COBOL compilation errors, we evaluate it on two types of LLMs: GPT variants (GPT-3.5, GPT-4, GPT-4o-mini, and GPT-4o \citep{gpt3.5,achiam2023gpt,gpt4o}) and the COBOL-specific models -  mAInframer (7B, 13B, 34B) \citep{noauthor_bloop_nodate}, fine-tuned from CodeLlama \citep{roziere2023code}. While our approach can be applicable to any instruction LLM, we focus on leading COBOL code models for reproducibility. GPT models, though not fine-tuned for COBOL, have demonstrated strong performance in general code generation and debugging tasks due to their broad training and instruction-following capabilities \citep{cipriano2024llms,liang2024can,soliman2025large}. 
% These models are trained on diverse and extensive datasets, enabling them to handle various coding languages and respond effectively to instruction-based prompts. Although they are not specifically fine-tuned for COBOL, their contextual understanding and reasoning make them useful tools for tasks such as automated debugging, where interpreting errors and improving code are critical. 
% To enhance the COBOL code generation capabilities, the mAInframer series was introduced as a family of fine-tuned models specific for COBOL, built on top of CodeLlama \citep{roziere2023code}. Compared to general LLMs, mAInframer models demonstrate significant improvements in both functional correctness and compilation rates. The performance metrics~\jinqiu{Add the exact numbers} for mAInframer-7B, 13B, and 34B,
In contrast, mAInframer is fine-tuned specifically on COBOL code and achieves higher compilation and functional correctness rates than GPT and other base models \citep{noauthor_bloop_nodate}. As shown in the COBOLEval benchmark \citep{noauthor_bloop_nodate}, mAInframer-34B achieved a pass@1 score of 10.27, surpassing GPT-4 and outperforming CodeLlama models.
% ~\jinqiu{Need to briefly describe how this family of models are fine-tuned from which general-purpose LLMs.}
% evaluated using COBOLEval, highlight their superiority over models like GPT models and CodeLlama. \citep{noauthor_bloop_nodate}.
% However, since the mAInframer series is not designed to function as an instruction model, it is unsuitable for the self-debugging process. For evaluating mAInframer models in our experiment, we only use mAInframer-34B to generate COBOL code and GPT-4o to perform the COBOLAssist process and debug the generated code.

 % Evaluations using COBOLEval demonstrate that state-of-the-art general-purpose LLMs, including GPT-4 \citep{gpt4} and CodeLlama \citep{roziere2023code}, struggle with COBOL code generation, achieving low pass@1 scores and compilation rates. This highlights the need for specialized models trained explicitly for COBOL.

\noindent\textbf{Compute Resources and Hyperparameters:} 
All experiments are executed on an NVIDIA H100 GPU machine with 80 GB of RAM to handle large model inference and code generation tasks.
Each program in the benchmark is compiled using the GnuCOBOL compiler (version 3.2.0) \citep{noauthor_bloop_nodate} to identify compilation errors, which serve as input for our COBOLAssist pipeline. To maintain consistency across evaluations, the chosen LLMs are queried via their respective APIs or local deployments under the same computational settings.
For a fair comparison, we set the temperature parameter to 0 to enforce deterministic outputs and minimize variability in code generation across repeated runs (similar to the experiments from BloopAI \citep{noauthor_bloop_nodate}). 
% Additionally, we limit the maximum number of self-debugging iterations (i.e., rounds of compilation feedback and regeneration) to 3. This constraint reflects a realistic balance between correction attempts and practical usability. An ablation study of this hyperparameter is discussed in more detail in Section \todo{label need to fixed}

\noindent\textbf{Metrics:}
We employ the following metrics to assess the performance of COBOLAssist:
\begin{itemize}
    \item \textbf{Compilation Success Rate (CSR)}: refers to the percentage of code samples that compile without errors when processed by a compiler.
    
    \item \textbf{Pass@k}: For each problem, \textit{k} program samples are generated; if at least one of the \textit{k} samples passes the unit tests, the task is considered successfully solved. The pass@k metric measures the overall percentage of solved tasks.
    % ~\jinqiu{ be specific `from k samples'}. 
    In our experiments, we report results using pass@1, evaluating only the top-1 program generated.
\end{itemize}

\subsection{Experiment Results}

\subsubsection{RQ1: What are the major categories of compilation errors in LLM-generated COBOL code?}
\label{rq1_results}
\begin{figure*}
    \centering
    \subfloat[\centering Human-written COBOL code (*)]{\includegraphics[width=6.5cm]{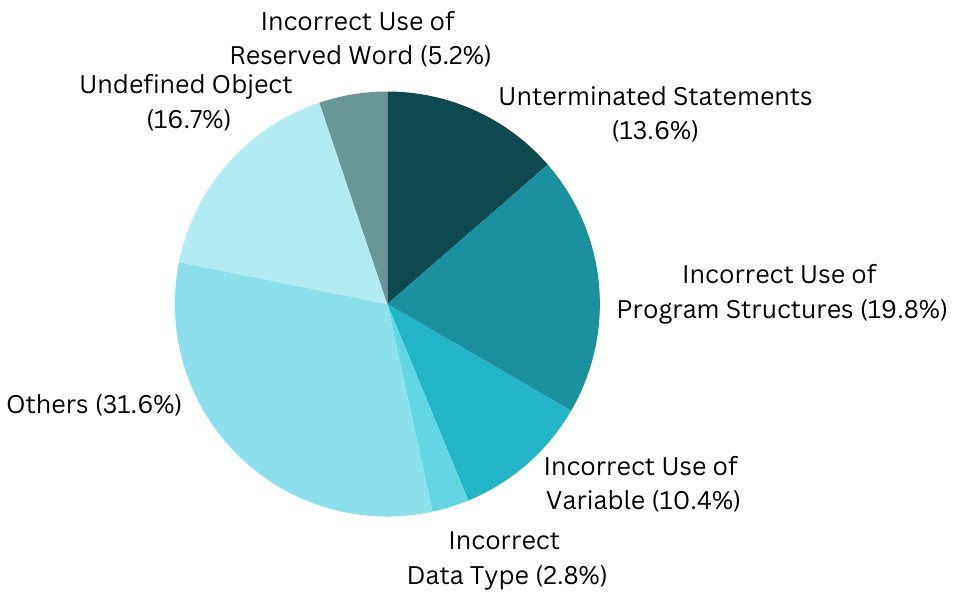}\label{subfig:human_written}}%
    \hfill
    \subfloat[\centering LLM-generated COBOL Code \\(Before Using COBOLAssist)]{\includegraphics[width=6cm]{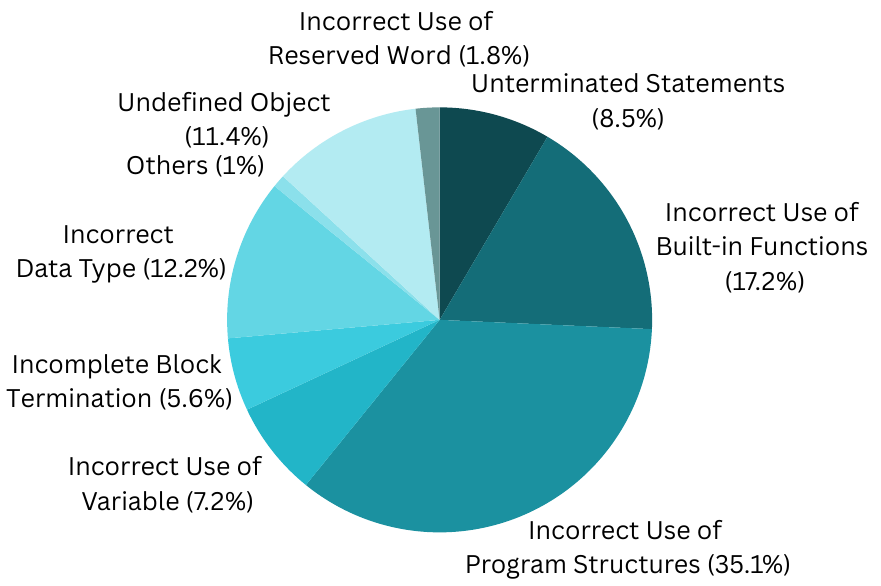} \label{subfig:before}}%
    \par
    \subfloat[\centering LLM-generated COBOL Code \\(After Using COBOLAssist)]{\includegraphics[width=6cm]{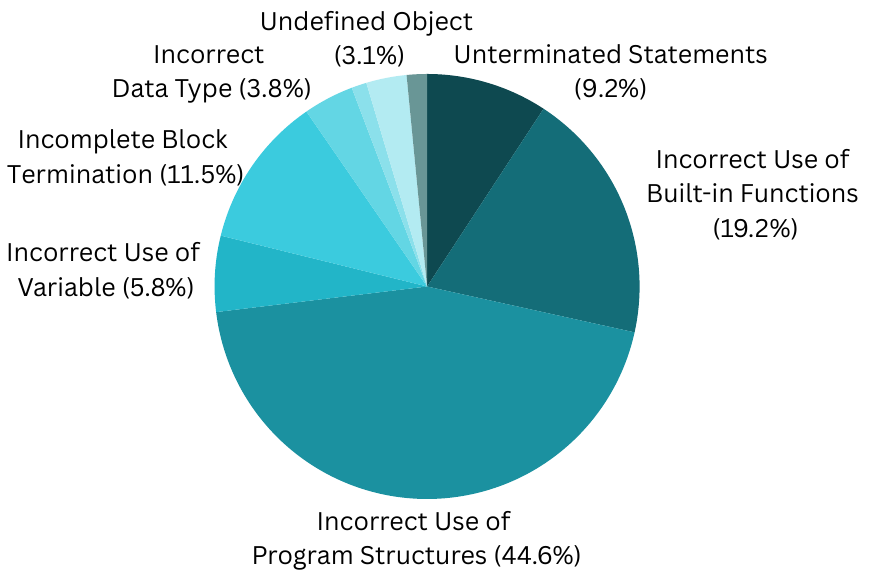} \label{subfig:after}}
    \caption{Compilation Error Categorization of Human-Written and LLM-generated COBOL Code (Before and After Using COBOLAssist). (*) The categories shown in this figure are grouped and renamed from the original categories presented in prior work \citep{litecky1974study} to align with our classification scheme. Details on the mapping process can be found in the artifact.}%
    \label{fig:example}
\end{figure*}

% \noindent \textbf{Method}
% \todo{need to update}
To better align the categories from our work and the prior study that investigates compilation errors in human-written code \citep{litecky1974study} for an easy comparison, we group similar error categories and rename them to align with our defined categories. We include details of the refinement and category mapping in our artifact. 
% Details of this step can be found in Section \ref{apd}
% \jinqiu{Should avoid forward reference, refer to the artifact ``We detail the refinement and category mapping in the artifact.''}. 

% Overall, both versions share several common error types, yet there are noticeable differences in their distribution and characteristics.
\noindent \textbf{Results}
Figure \ref{subfig:human_written} and \ref{subfig:before} illustrate the distribution of compilation error types in human-written COBOL code and LLM-generated COBOL code before applying COBOLAssist.
While both human and generated code share common error types, notable differences emerge in their distribution patterns.
Notably, \textit{Incorrect Use of  Program Structures} accounts for the largest proportion of errors in both human-written and LLM-generated code, but it is more severe in generated code (35.1\%) than in human-written code (19.8\%). This gap indicates that while LLMs grasp COBOL’s basic syntax, they are still not good at mastering its complex structured programming rules. Moreover, two new error categories emerge in the LLM-generated code: \textit{Incorrect Use of Built-in Functions} (17.2\%) and \textit{Incomplete Block Termination} (5.6\%), which are absent in human-written code. These errors are likely caused by the model’s limited understanding of COBOL’s built-in function usage and structural conventions, which are typically better handled by human developers.

\begin{tcolorbox}
 \label{finding1}
  \textbf{Finding 1:} LLMs show a relatively high rate of errors in structural usages (35.1\%) compared to human-written COBOL code, underscoring their struggle with COBOL’s structural logic.
\end{tcolorbox}

\begin{tcolorbox}
\label{finding2}
\textbf{Finding 2:} Two new error types (i.e., \textit{Incorrect Use of Built-in Functions} and \textit{Incomplete Block Termination}) are unique to LLM-generated COBOL code, which highlights weaknesses of LLMs in handling COBOL-specific semantics.
\end{tcolorbox}

In contrast, human-written code has a higher percentage of \textit{Others} errors (31.6\%) compared to generated code (1\%), indicating that human errors tend to be more diverse and often context-specific. For other shared categories such as \textit{Incorrect Use of Variable, Incorrect Data Type}, and \textit{Unterminated Statements}, the proportions are roughly comparable across both code sets. These findings show that LLMs might achieve near-human proficiency in handling syntactic elements of COBOL.
% that while LLMs are capable of generating COBOL code with an error profile somewhat similar to human developers in most categories, they are particularly prone to mistakes in the use of control structures and functions — two critical aspects of COBOL programming that require further improvement.

\begin{tcolorbox}
\label{finding3}
\textbf{Finding 3:} Human-written COBOL code has broader and more diverse compilation errors, while LLM-generated errors are more concentrated with limited types.
\end{tcolorbox}

\begin{tcolorbox}
\label{finding4}
\textbf{Finding 4:} 
% For basic syntax and type-related categories (e.g., \textit{Incorrect Use of Variables} and \textit{Incorrect Data Types}), LLMs perform comparably with human developers.
LLMs share similar error patterns with human developers in basic syntax and type-related categories (e.g., \textit{Incorrect Use of Variable} and \textit{Incorrect Data Type}), suggesting that they are reasonably capable of handling fundamental COBOL constructs.
%\jinqiu{Revise the tone a bit, as this is not about competition.. ``LLMs ''}
\end{tcolorbox}

\subsubsection{RQ2: How effective is COBOLAssist in resolving COBOL compilation errors?}

\begin{table}[ht]
\renewcommand{\arraystretch}{1.2}
\centering
\begin{tabular}{c|c|c|c}
\hline
Model & Setting & CSR (\%) & pass@1 \\ \hline

\multirow{3}{*}{GPT-3.5} 
& w/o COBOLAssist 
& 13.72 & 3.62 \\
& Zero-shot refinement (no compiler) 
& 15.07 & 4.11 \\
& w/ COBOLAssist 
& 21.38 & 7.13 \\ \hline

\multirow{3}{*}{GPT-4} 
& w/o COBOLAssist 
& 31.91 & 9.10 \\
& Zero-shot refinement (no compiler) 
& 37.67 & 13.70 \\
& w/ COBOLAssist 
& 55.17 & 22.60 \\ \hline

\multirow{3}{*}{GPT-4o-mini} 
& w/o COBOLAssist 
& 29.50 & 8.22 \\
& Zero-shot refinement (no compiler) 
& 43.15 & 10.27 \\
& w/ COBOLAssist 
& 64.38 & 15.07 \\ \hline

\multirow{3}{*}{GPT-4o} 
& w/o COBOLAssist 
& 41.80 & 16.40 \\
& Zero-shot refinement (no compiler) 
& 54.79 & 20.54 \\
& w/ COBOLAssist 
& 95.89 & 29.45 \\ \hline

\multirow{3}{*}{mAInframer-34B} 
& w/o COBOLAssist 
& 74.66 & 10.27 \\
& Zero-shot refinement (no compiler) 
& 74.66 & 10.27 \\
& w/ COBOLAssist (via GPT-4o) 
& 97.94 & 11.67 \\ \hline

\end{tabular}
\caption{Effectiveness of COBOLAssist across different LLMs on COBOLEval. Zero-shot refinement refers to iterative self-correction without compiler feedback.}
\label{tab:compile_success_rate}
\end{table}

\noindent \textbf{Method.}
To answer this research question, we report the CSR and pass@1 for five models: GPT-3.5, GPT-4, GPT-4o-mini, GPT-4o, and mAInframer-34B—under three settings: (1) without COBOLAssist, (2) zero-shot self-refinement without compiler feedback, and (3) with COBOLAssist, as shown in Table ~\ref{tab:compile_success_rate}. 
The zero-shot refinement setting iteratively prompts the model to improve its own outputs without access to compiler feedback, allowing us to isolate the contribution of compiler-guided debugging. For COBOLAssist, we limit the maximum number of self-debugging iterations (i.e., rounds of compilation feedback and regeneration) to 3, reflecting a practical balance between effectiveness and computational cost. We further analyze the impact of this threshold in the ablation study.
Since mAInframer-34B is not designed as an instruction-following model, it is not suitable for iterative self-debugging. Therefore, in the COBOLAssist setting, we use it only for initial COBOL code generation, while GPT-4o performs the debugging process. 
% We also evaluate a zero-shot refinement setting using mAInframer-34B alone; however, this setting does not yield any improvement.

\noindent \textbf{Results.}
Our experiments show that COBOLAssist significantly improves the compilation success rate across all evaluated models, outperforming both the baseline and zero-shot refinement settings. While zero-shot refinement provides modest gains (e.g., GPT-4 improves from 31.91\% to 37.67\%, and GPT-4o from 41.80\% to 54.79\%), these improvements remain limited compared to compiler-guided debugging.
With COBOLAssist, GPT-3.5 improves from 13.72\% to 21.38\%, while GPT-4 increases from 31.91\% to 55.17\%. GPT-4o-mini shows a substantial gain from 29.50\% to 64.38\%, and GPT-4o achieves the most significant improvement, reaching 95.89\% compared to its initial 41.80\%. For mAInframer-34B, zero-shot refinement does not yield any improvement, with performance remaining at 74.66\%. In contrast, COBOLAssist further increases its performance to 97.94\% when paired with GPT-4o.

% Before debugging, GPT-3.5 achieves the lowest success rate at 13.72\%; GPT-4o-mini performs better, reaching 29.5\%, followed by GPT-4o-mini (31.9\%) and GPT-4o (41.8\%); mAInframer-34b stands out, achieving a 74.66\% success rate due to its COBOL specialization. After the self-debugging process, all models demonstrate significant improvements. GPT-3.5 increases to 21.38\%, while GPT-4 nearly doubles its rate to 55.17\%. GPT-4o-mini improves to 64.38\%, and GPT-4o demonstrates a remarkable leap to 95.89\%. mAInframer-34b, when combined with GPT-4o for debugging, reaches 97.94\%. 

These results highlight the critical role of compiler feedback in enabling reliable error correction, particularly for models with strong instruction and reasoning capabilities. The large gap between zero-shot refinement and COBOLAssist demonstrates that iterative prompting alone is insufficient for resolving the  syntactic constraints of COBOL.

% \begin{tcolorbox}
%  \label{finding1}
%   \textbf{Finding 5:} LLMs can effectively fix COBOL compilation errors by using the compiler feedback. \jinqiu{make this one about summarizing all the results; well simplify finding 8 and swap with this one.}
% \end{tcolorbox}

% \begin{tcolorbox}
%  \label{finding1}
%   \textbf{Finding 6:} The ability to fix compilation bugs depends heavily on the chosen LLM. \jinqiu{this lacks context and too general.}
%   % Models like GPT-4o outperform others, indicating that both general reasoning ability and domain specialization play key roles.
% \end{tcolorbox}
   
% Following the improvement in compile success rate, we also investigate the effectiveness of COBOLAssist by reporting the pass@1 score, which measures the correctness of the generated code on the first attempt. GPT-3.5 and GPT-4 show improvements from 3.62 to 7.13 and 9.1 to 22.6, respectively. GPT-4o-mini improves from 8.22 to 15.07, and GPT-4o achieves the highest jump, from 16.4 to 29.45. On the other hand, while mAInframer achieves a relatively high compile success rate (as shown in Table \ref{tab:compile_success_rate}), its pass@1 only increases slightly from 10.27 to 11.67. This suggests that while its code is compilable, it may often be semantically incorrect or incomplete. 

We further evaluate functional correctness using pass@1. Consistent with CSR improvements, COBOLAssist improves pass@1 across all models. GPT-3.5 and GPT-4 show improvements from 3.62 to 7.13 and 9.1 to 22.6, respectively. GPT-4o-mini improves from 8.22 to 15.07, and GPT-4o achieves the highest jump, from 16.4 to 29.45. Zero-shot refinement again yields smaller gains, reinforcing the importance of compiler feedback.
In contrast, mAInframer-34B shows only a marginal improvement in pass@1 (10.27 to 11.67) despite its high compilation success rate. This indicates that while the generated code is often syntactically valid, it frequently remains semantically incorrect or incomplete. This gap between compilation success and functional correctness highlights a key limitation of current approaches and motivates future work on semantic-aware repair strategies.

% \begin{tcolorbox}
%  \label{finding1}
%   \textbf{Finding 7:} COBOLAssist can enhance not only compilation but also functional correctness.\jinqiu{this is simple should be merged with finding 8.}
% \end{tcolorbox}

\begin{tcolorbox}
 \label{finding5}
  % \textbf{Finding 5:} COBOLAssist improves both the compilation success rate and functional correctness across all evaluated LLMs. GPT-4o demonstrates the best overall performance among the GPT series, achieving the highest compile success rate (95.89\%) and pass@1 score (29.45\%). While mAInframer-34B reaches the highest compilation success rate (97.94\%) due to its COBOL-specific training, its lower Pass@1 score (11.67\%) exhibits a non-trivial gap in functional correctness compared to state-of-the-art general-purpose LLMs, indicating that syntactic correctness does not always imply semantic accuracy.
    \textbf{Finding 5:} COBOLAssist improves both compilation success rate and functional correctness across all evaluated LLMs. Among the GPT series, GPT-4o achieves the best overall performance, with a compilation success rate of 95.89\% and a pass@1 score of 29.45\%. Although mAInframer-34B attains the highest compilation success rate (97.94\%) due to its COBOL-specific training, its lower pass@1 score (11.67) indicates a substantial gap in functional correctness compared to state-of-the-art general-purpose LLMs. 
\end{tcolorbox}
         
\begin{table}[ht]
\renewcommand{\arraystretch}{1.5}
\centering
\begin{tabular}{c|r|c|c}
\toprule
Model& Before $\rightarrow$After & Compilation Error &Avg. Iteration \\
& COBOLAssist & Reduction (\%$\downarrow$) & Number\\ \hline
GPT-3.5        & 487 $\rightarrow$ 325       & 33.70 & 1.79 \\ 
GPT-4          & 325 $\rightarrow$ 182    & 44.00 &0.98 \\                                                                
GPT-4o-mini    & 387 $\rightarrow$  201    & 48.06 &1.53 \\             
GPT-4o         & 243 $\rightarrow$ 46   & 81.07 &1.39 \\                                                                              
mAInframer-34B (with GPT-4o) & 90 $\rightarrow$ 13   & 85.56 &1.56 \\                            \bottomrule
\end{tabular}
\caption{Effectiveness of COBOLAssist in reducing compilation errors across different models. The table reports the number of compilation errors before and after applying COBOLAssist, the corresponding percentage reduction, and the average number of iterations required for correction. For mAInframer-34B, COBOLAssist is performed using GPT-4o for the debugging process.}
\label{tab:avg_round}
\end{table}

% \jinqiu{What is the threshold of iteration? Unlimited?}\todo{this is not a hyperparameter, put it under \noindent\textbf{Method} of this RQ.}
% \anh{I consider it as a hyperparameter and set to 3 (explained in section 'Compute Resources and Hyperparameters'), more configurations can be found in ablation study.}
To further provide insights into the effectiveness and efficiency of COBOLAssist, we report the number of compilation errors before and after applying COBOLAssist, the corresponding percentage reduction in compilation errors, and the average number of iterations required by COBOLAssist to reach a compilable program in Table \ref{tab:avg_round}. As shown in the table,  COBOLAssist consistently reduces compilation errors while requiring fewer than two iterations on average.
\begin{figure*}[]
\centering
\includegraphics[width=0.8\textwidth]{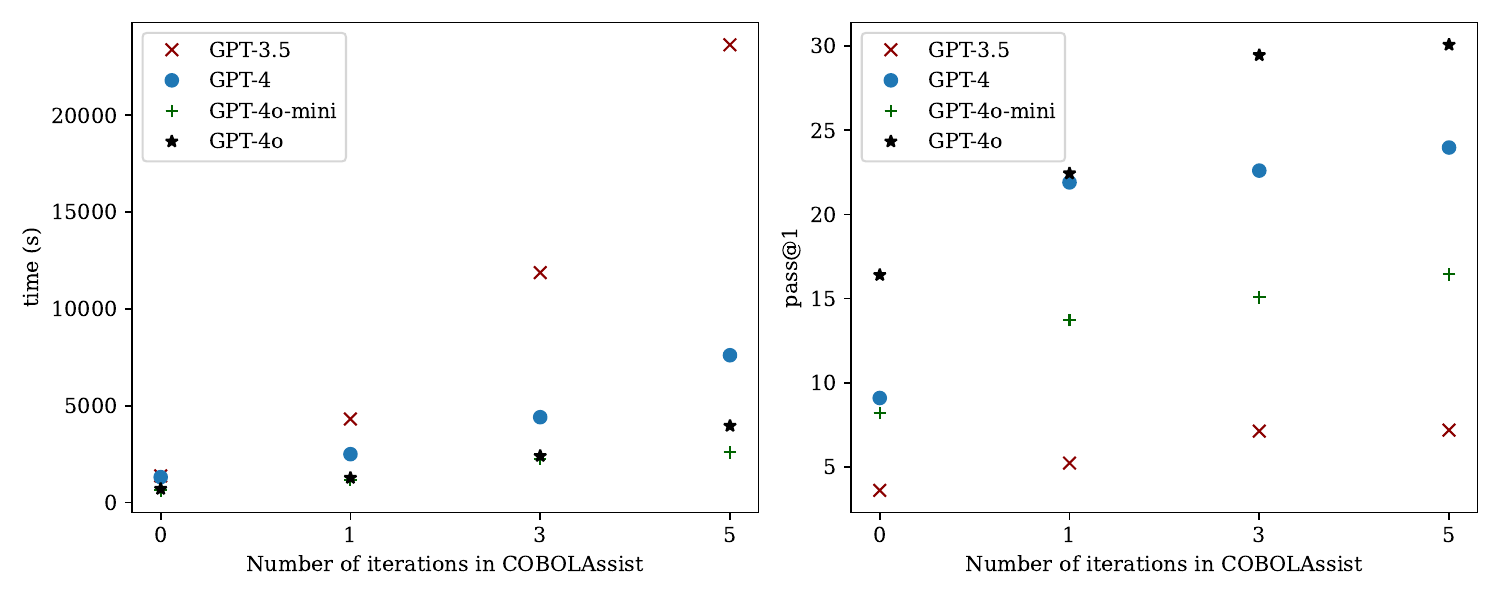}
\caption{Impact of COBOLAssist iteration configuration on execution time and pass@1 across different models.}
\label{fig:multi_LLM}
\end{figure*}
Among the evaluated models, mAInframer-34B achieves the highest error reduction rate of 85.56\%, lowering the number of compilation errors from 90 to 13 in just 1.56 iterations on average. GPT-4o follows closely, reducing errors by 81.07\% (from 243 to 46) with an average of 1.39 iterations, making it the most efficient model in terms of iteration count. GPT-4o-mini and GPT-4 achieve error reductions of 48.06\% and 44.00\%, respectively, showing substantial improvements over their original states. In contrast, GPT-3.5 is the least effective, reducing only 33.70\% of its compilation errors and requiring the most iterations (1.79 on average).

% Meanwhile, mAInframer requires 1.56 rounds on average and reduces errors from 90 to 13. 

\noindent \textbf{Ablation study on the number of iterations allowed}

We conduct an ablation study to evaluate the effect of the number of iterations on the performance of COBOLAssist. We experiment with four configurations, where the number of allowed iterations is set to 0, 1, 3, and 5, respectively. A value of 0 indicates that COBOLAssist is not applied—only the raw output from the base LLM is considered. We analyze how this parameter affects both effectiveness (measured by pass@1) and execution time (total time spent in seconds on generation and verification).

Figure \ref{fig:multi_LLM} presents the results with four different models: GPT-3.5, GPT-4, GPT-4o-mini, and GPT-4o. Across all LLMs, pass@1 steadily improves with more iterations. GPT-4o shows the strongest gains, improving from 16.4 (0 iterations) to 30.07 (5 iterations), while maintaining a relatively modest runtime increase. GPT-4 also benefits significantly, reaching 23.97 pass@1 at 5 iterations. In contrast, GPT-3.5 starts with a much lower baseline (3.62) and improves only marginally after five iterations (7.19), despite a sharp increase in total time from 1,380s to 23,637s. This suggests that weaker models (e.g., GPT-3.5) are less effective in fixing compilation errors, even after trying several iterations of fixing. %require more iterations to fix compilation errors but still lacking in effectiveness.
% \begin{figure*}[h]
% \centering
% \includegraphics[width=0.9\textwidth]{image/flow.pdf}
% \caption{
% A Sankey diagram shows the transformation of compilation error categories before and after applying COBOLAssist. The left and right sides show the compilation errors before and after using COBOLAssist. \textit{ERRORS HAVE BEEN FIXED} indicates successful repairs by COBOLAssist, and \textit{INTRODUCING NEW ERRORS} captures new compilation errors.}
% \label{fig:sankey}
% \end{figure*}
These results indicate that COBOLAssist significantly boosts the performance of all models, particularly when paired with strong LLMs like GPT-4 and GPT-4o. Furthermore, increasing the number of iterations beyond three provides diminishing returns for most models, especially the weaker ones. In practice, a configuration with three iterations strikes a good balance between runtime and effectiveness.

\subsubsection{RQ3: What types of compilation errors remain unresolved after using COBOLAssist?}

\noindent \textbf{Results}
To further understand the limitations of the LLM-based self-debugging approach, we analyze and compare the distribution of compilation error types in the generated COBOL code before and after using COBOLAssist, as shown in figures \ref{subfig:before} and \ref{subfig:after}. 

We observed significant reductions in two major error categories: \textit{Incorrect Data Type} and \textit{Undefined Object}. The number of \textit{Incorrect Data Type} errors dropped from 12.2\% to 3.8\%, while \textit{Undefined Object} errors decreased from 11.4\% to 3.1\%.
% \jinqiu{need to comment whether they turn into other types of compilation errors as side effects from LLM repairs.} 
These results indicate that LLMs are particularly effective in resolving such errors, as the compiler log provides both descriptive error messages and exact line references. This information enables the model to pinpoint and correct these errors. Consequently, most bugs in these categories can be resolved without introducing new types of compilation errors as side effects from LLM repairs.
% These results indicate that LLMs are particularly effective in resolving these types of errors, which are often straightforward and can be corrected based on error messages and context understanding.

% \jinqiu{I don't find this type in the figure, just comment on the change and flow? E.g., half is fixed, yet many remain and many turn into another category, i.e., LLMs introduce xxx errors after fixing these errors.} \anh{images have been updated. i didnt trigger if fixing one errors may turn into another one, but the new errors introduced after debugging are shown in sankey diagram}
% \jinqiu{we should not compare the percentage here, maybe just comment on which categories are top now? top 3 changes etc.?}
% However, the proportion of syntax-related errors, especially those involving\textit{ Incorrect Use of Control Structures}, \textit{Incorrect Use of Built-in Functions}, and \textit{Incomplete Block Termination}, increases after the self-debugging process. For instance, \textit{Incorrect Use of Control Structures} rises from 35.1\% to 44.6\%, \textit{Incorrect Use of Built-in Functions} grows from 17.2\% to 19.2\%, and \textit{Incomplete Block Termination} increases from 5.6\% to 11.5\%. This trend suggests that while the LLM can address simple semantic or naming issues, it struggles with correcting complex structural or syntax-related mistakes in COBOL code. Moreover, \textit{Unterminated Statements} remain nearly unchanged (from 8.5\% to 9.2\%), further highlighting the challenge LLMs face in properly handling COBOL’s syntax rules.
After applying COBOLAssist, the top three remaining compilation error categories are \textit{Incorrect Use of  Program Structures}, \textit{Incorrect Use of Built-in Functions}, and \textit{Incomplete Block Termination}.  Despite the improvements seen in overall compilation success, these error types often require deeper contextual understanding and multi-line reasoning—skills that current LLMs still struggle with. 
For example, LLMs frequently fail to insert required terminators such as \texttt{END-IF}, especially in nested control structures that span multiple lines. These cases demand the model correctly track and align opening and closing constructs across extended code blocks. When such terminators are omitted, the resulting code is structurally invalid and fails to compile.
This pattern highlights that LLMs are still limited in addressing complex syntax and structural rules inherent to COBOL.

\begin{tcolorbox}
 \label{finding6}
  \textbf{Finding 6:} When provided with compiler feedback, LLMs are effective at repairing errors related to \textit{Incorrect Data Type} and \textit{Undefined Object}.
\end{tcolorbox}

\begin{tcolorbox}
 \label{finding7}
  \textbf{Finding 7:} LLMs struggle to fix the bug related to the COBOL syntax, particularly those involving  program structures, function usage, and block termination, likely due to the structural complexity and syntax requirements of the COBOL programming language.
\end{tcolorbox}

\section{Threats to Validity}
% Like any empirical study, there are threats to the validity of our findings. We discuss below the most significant ones of these, as well as the mitigating factors.

\noindent\textbf{Internal Threats.}
% One potential threat to internal validity stems from the inherent nondeterminism of large language models. Although we set the temperature to 0 to minimize output variability, the same prompt may still yield slightly different outputs across different runs. To account for this, we executed each experiment three times and reported the average results to reduce the impact of output variance on our findings. Another internal validity threat arises from our manual analysis process, particularly during the categorization of compilation errors. Although we designed a taxonomy tailored to COBOL and refined it iteratively during labeling, human subjectivity may still influence how certain errors are interpreted or grouped. 
% To mitigate this threat, two annotators independently reviewed and resolved disagreements after each labeling round, and one author verified the final set to ensure consistency.
LLMs are inherently nondeterministic. Although we set the temperature to 0, minor output variations may still occur. To mitigate this, we ran each experiment three times and reported average results. Another threat arises from manual error categorization, which may introduce subjectivity. To reduce bias, two annotators independently labeled errors and resolved disagreements, and one author verified the final taxonomy.

\noindent\textbf{External Threats.}
Our evaluation is limited to the COBOLEval dataset (146 tasks), the only publicly available benchmark for COBOL code generation. While it provides diverse tasks, results may not generalize to larger industrial COBOL systems. Additionally, we evaluated a limited set of LLMs. Broader datasets, models, and real-world codebases should be explored in future work.
% Threats to external validity concern the extent to which our findings can be generalized beyond the specific experimental settings used in this study. A key limitation is that we evaluated our self-debugging approach solely on the COBOLEval dataset, which consists of 146 COBOL programming tasks. We selected this dataset because it is currently the only publicly available benchmark specifically designed for COBOL code generation. While COBOLEval provides a diverse set of tasks and serves as a reasonable starting point for evaluating COBOL-related models, the generalizability of our findings to other COBOL codebases or real-world legacy systems remains to be further explored. Additionally, our study focused only on COBOL programming and a limited set of LLMs. As COBOL differs significantly in syntax and structure from modern programming languages, further investigation is needed to assess whether the proposed self-debugging strategy can be adapted effectively to other languages or to more complex code generation tasks, such as multi-function programs or full-system generation. Future work could extend the evaluation to larger-scale datasets, industrial COBOL applications, or a broader range of LLMs and prompting strategies to validate the robustness and applicability of our findings across various real-world settings.

\noindent\textbf{Construct Validity.}
We rely on GnuCOBOL compiler outputs to identify compilation errors. Although compilation success is a practical proxy for syntactic correctness, it does not capture deeper semantic or logical errors. Future work should incorporate semantic evaluations, such as test-based validation or expert review, to provide a more comprehensive assessment of code quality.

% Construct validity refers to whether the metrics and experimental design accurately capture what we intend to measure—in this case, the correctness and characteristics of compilation errors in LLM-generated COBOL code. Our study primarily relies on the output of the GnuCOBOL compiler to detect and categorize errors, as recommended by the original COBOLEval benchmark \citep{noauthor_bloop_nodate}. While compilation success or failure is a practical and widely accepted proxy for assessing syntactic and some semantic correctness, it does not reflect deeper logical or functional correctness. As a result, programs that compile successfully may still contain subtle bugs or incorrect logic that the compiler cannot detect. Another threat to construct validity lies in our focus on compilation errors, which primarily reflect syntactic and some structural issues in the code. This approach does not capture deeper semantic or logical errors, which may still exist even in code that compiles successfully. Future work should incorporate semantic-level evaluations, such as test-based validation or expert review, to provide a more comprehensive understanding of code correctness.

\section{Related Work}
\subsection{LLM-Generated Code Analysis and Bugs
}
Several studies have empirically analyzed the correctness and quality of LLM-generated code~\citep{fan2023automated,tambon2024bugs,lian2024uncovering,liu2024your,pan2024lost}. LLMDefects \citep{fan2023automated} contains mistakes made by Codex-generated Java code. A recent study \citep{tambon2024bugs} examines the characteristics of bugs in AI-generated code, while \citep{lian2024uncovering} evaluate multiple code models to assess their quality and weaknesses. \citep{liu2024your} introduce \textit{EvalPlus}, a framework to benchmark functional correctness in LLM-synthesized code, while another study \citep{liu2024refining} analyzes correctness, quality issues, and the self-repair capabilities of ChatGPT. Additionally, \citep{pan2024lost} investigate LLM performance in code translation, exploring error patterns and potential improvements through prompting techniques. While these studies provide insights into various error aspects of generated code, they do not focus on compilation errors in COBOL, which we aim to address in our research.

\subsection{Automated Program Repair with LLMs
}
To improve the quality of the LLM-generated code, several studies have proposed automated program repair (APR) approaches. \citep{fan2023automated} propose using APR techniques to fix mistakes in LLM-generated code.  \citep{jin2023inferfix} present an end-to-end framework that combines transformers with static analysis to address security and performance bugs.  \citep{chen2023teaching} introduce a methodology in which LLMs learn to debug their own generated code using few-shot learning. Additionally, \citep{connor2022can} explore LLM-driven edit operations for fixing incorrect solutions.

For syntax error correction, \citep{ahmed2022synshine} propose a deep-learning-based tool for fixing syntactic mistakes, while Zhu et al. \citep{zhu2021syntax} present \textit{Recorder}, an edit-generation model that ensures syntactic correctness. Notably, \citep{deligiannis2024rustassistant} focus on automatically fixing Rust compilation errors using LLMs. However, the effectiveness of such techniques depends heavily on the underlying model’s capabilities, which may change as LLMs evolve. Our work extends this direction by focusing on COBOL compilation errors, which pose unique challenges due to legacy system constraints.

\subsection{Compiler Error Handling and Debugging}
The role of compiler error messages in developer productivity and automated debugging has been extensively studied. \citep{becker2019compiler} and  \citep{traver2010compiler} analyze the effectiveness of compiler messages in software development.  \citep{litecky1974study} investigates error frequencies in COBOL, while \citep{ciborowska2021contemporary} highlights the distinct challenges in mainframe maintenance. Although these studies provide valuable insights into compiler error handling and COBOL-specific defect patterns, they do not explore how LLMs contribute to mitigating COBOL compilation failures. Our research aims to bridge this gap by systematically analyzing common compilation errors in LLM-generated COBOL code and proposing automated strategies to improve compilation success rates.

\section{Discussion and Implications}
Based on our study and our evaluation, we discussed the implications for developers using LLM-based code generation models and researchers of code generation models.

\noindent\textbf{Implications for Developers.} Our categorization of compilation errors in LLM-generated code lays the first step in understanding the errors in LLM-generated code. As our study shows that LLM-generated code may exhibit different types of compilation errors compared to human-written COBOL code (Findings 1 and 2), 
developers who would like to use publicly available LLMs (e.g., GPT-4o-mini) to support the maintenance of COBOL code need to be aware of these error types during LLM-assisted coding, especially the new error types (e.g., Incorrect Use of Built-in Functions), as they are uncommon in human-written code. As our proposed workflow of COBOLAssist that uses compilation feedback to guide LLMs to repair the compilation errors is general, we foresee that developers can use a similar workflow to reduce the time and effort when maintaining COBOL code. 

\noindent\textbf{Implications for Researchers.}
Finding 2 of our study revealed two new error types (i.e., \textit{Incorrect Use of Built-in Functions} and \textit{Incomplete Block Termination}). Both types show the inability of LLMs to effectively understand
the syntax of COBOL code (particularly due to its structured programming approach). Although COBOLAssist manages to improve on the compilation success rate of existing LLMs, we foresee that a syntax-aware approach that incorporates program analysis techniques (e.g., abstract syntax tree) to better capture the structured programming approach can further help to improve the results. Moreover, as LLMs share similar error patterns with human developers (Finding 4) and there is no specialized automated tool available for fixing COBOL code, our study also provides empirical evidence showing the feasibility of using LLMs to fix compilation errors in human-written code.

\section{Conclusion}
In this paper, we conducted an empirical study to investigate the capability of LLMs in generating and debugging COBOL code, a crucial but underexplored domain. Our study identifies unique characteristics of compilation errors in LLM-generated COBOL code. We proposed a self-debugging framework named COBOLAssist to iteratively identify and fix COBOL compilation errors using compiler feedback. Our experiments demonstrated that our approach improves the compilation success rate and correctness of generated COBOL programs across different LLMs. From our findings, this empirical study sheds light on the challenges of COBOL code generation and highlights the potential of LLMs combined with self-debugging techniques to support legacy code modernization.

\begin{acknowledgements}
The authors gratefully acknowledge Nguyen Le Phuoc, a COBOL expert from FPT Software AI Center for his support in reviewing and contributing to the development of the error taxonomy.
\end{acknowledgements}

\section*{Declarations}

\noindent\textbf{Funding:}  
Not applicable.

\noindent\textbf{Ethical Approval:}  
Not applicable.

\noindent\textbf{Informed Consent:}  
Not applicable.

\noindent\textbf{Author Contributions:}  
Conceptualization: Anh T. V. Dau, Shin Hwei Tan, Jinqiu Yang, Nghi D. Q. Bui, Anh Tuan Nguyen.
Data Curation: Anh T. V. Dau. 
Methodology: Anh T. V. Dau, Shin Hwei Tan, Jinqiu Yang, Anh Tuan Nguyen. 
Experiment Analysis: Anh T. V. Dau, Shin Hwei Tan, Jinqiu Yang, Anh Tuan Nguyen.
Supervision, supervising experiment design and execution: Shin Hwei Tan, Jinqiu Yang, Anh Tuan Nguyen.
Writing – Original Draft: Anh T. V. Dau, Shin Hwei Tan, Jinqiu Yang.
Writing – Review \& Editing: Anh T. V. Dau, Shin Hwei Tan, Jinqiu Yang

\noindent\textbf{Data and Code Availability Statement:}  
The artifact, data and code used in this project are publicly available:  \url{https://github.com/anonymizedName/COBOLAssist}.

\noindent\textbf{Conflict of Interest:}  
The authors declare that they have no conflict of interest.

\noindent\textbf{Clinical Trial Number:}  
Not applicable.

\bibliographystyle{spbasic}     
\bibliography{reference}

\end{document}